\documentclass{article}
\usepackage{amssymb}
\usepackage{amsmath}
\usepackage{amsfonts}
\usepackage{sw20ssur}
\usepackage{cite}
\usepackage{afterpage}

\input{tcilatex}
\begin{document}

\title{Game theoretical modelling of network/cybersecurity}
\author{Azhar Iqbal$^{\dag \wr \curlyvee }$, Lachlan J. Gunn$^{\lozenge }$,
Mingyu Guo$^{\ddag }$,\textbf{\ }M. Ali Babar$^{\ddag }$, and Derek Abbott$%
^{\dag }$ \\
$^{\dag }${\small School of Electrical \& Electronic Engineering, University
of Adelaide, SA 5005, Australia.}\\
$^{\ddag }${\small School of Computer Science, University of Adelaide, SA
5005, Australia.}\\
$^{\lozenge }${\small Department of Computer Science, Aalto University,
Espoo FI-00076, Finland.}\\
$^{\wr }${\small Sage International Australia, PO Box 354, Torrensville SA
5031, Australia.}\\
$^{\curlyvee }${\small Interactive Decisions, PO Box 1025, Flinders Park SA
5025, Australia.}}
\maketitle

\begin{abstract}
Game theory is an established branch of mathematics that offers a rich set
of mathematical tools for multi-person strategic decision making that can be
used to model the interactions of decision makers in security problems who
compete for limited and shared resources. This article presents a review of
the literature in the area of game theoretical modelling of
network/cybersecurity.
\end{abstract}

\section{Introduction}

For most physical security situations the outcomes depend on the actions of
both attackers and defenders. The attackers and defenders act rationally and
can depend on various incentives that may be diametrically opposite or,
under other circumstances, may have some overlap. Physical security thus
provides the situations where the tools of game theory can be beneficially
applied and can provide insights into making optimal security decisions. For
various decision-making problems arising in physical security, game theory
can provide a rich set of analytical methods and mathematical tools.

The pervasive use of the Internet opens up numerous network security
situations. The attackers and defenders in typical situations are rational
agents who have the ability to act strategically. The agents can be assumed
to be interested in finding either the most damaging or the most secure use
of available resources. In the domain of network security, game theory has
been shown \cite{Liang2012,Roy2010} to provide useful insights in making
decisions that lead to developing novel, analytic, computational, and
practical approaches in the thought, policy, planning, and strategic action.
Game theory provides methodical approaches in order to explain the
inter-dependencies of the role of hidden and asymmetric information in
networks, network security decisions \cite{AlpcanBasar2010,Xiaolin2008}, the
incentives/limitations of the attackers, the perception of risks and costs
in human behavior, and much more. The elegant and powerful tools made
available by game theory are found to be highly useful in order to build
secure, resilient, and dependable networked systems \cite%
{Alpcan2003,Abdalzaher2016}.

\section{Game theory}

Game theory \cite%
{Myerson1991,Fudenberg1991,Harsanyi1988,Rasmusen1989,Gibbons1992,Osborne1994,Osborne2004}
is an established branch of mathematics that develops mathematical models,
allowing rigorous analysis, of strategic interaction between rational
decision-makers. It studies complex, competitive, and multi-agent
interactions in which one player's utility depends not only on his
decisions, but also on the decisions of his opponents. Game theory is
applied to a number of disciplines, including economics \cite{Neumann1944,
Aumann1995}, biology \cite{MaynardSmith1982,Weibull1997,Broom2013},
political science \cite{Morrow1994,Brams2011}, electrical engineering \cite%
{Bauso2016,MacKenzie2006}, business \cite{Papayoanou2010,Chatterjee2014},
computer science \cite{Krzysztof2011,Nisan2011}, law \cite%
{Baird1998,Iqbal2017}, public policy \cite{McCain2009}, physical security 
\cite{Rass2018,GameSec}, mechanism design \cite%
{Mingyu2018,Mingyu2017,Mingyu2016}, and more recently to the quantum
information \cite{ScholarGoogle2018,Iqbal20181,Iqbal20182,Khan2018}. Game
theory can enable opponents to predict each other's rational behavior and
suggest a course of action to be taken in any given situation.

\subsection{Static and dynamic games}

Games considered in the field of game theory are broadly classified as being
the \textit{static} or \textit{dynamic} games \cite{Rasmusen1989,Gibbons1992}%
. In static games the players choose their strategies simultaneously whereas
dynamic games involve a sequence of moves. In dynamic games \cite%
{Haurie2005,Basar2018}, a player chooses before others do, knowing that the
others' choices will be influenced by his/her publicly observable choice.
The dynamic character of the game results in models that can enhance the
learning ability of the players. In turn, this learning can help security
practitioners to develop high quality theoretical studies on real-life
problems. Depending on a particular situation in cybersecurity that are
amenable to game theoretical analyses, applying either static or dynamic
game theory can considered appropriate. For instance, a cybersecurity
situation involving a team of attackers and a plan of attack in which the
attackers act simultaneously, the application of static game theory will be
required. Whereas, dynamic game theory will be applied when some of the
attackers act first and the reaction of the defenders is observed before the
remaining attackers act while equipped with the knowledge of how the
defenders have reacted.

\subsection{Nash equilibrium}

The rule for predicting how a game will be played defines the \textit{%
solution concepts} in terms of which the game is understood by game
theorists. The most commonly used solution concept in game theory is that of
a Nash equilibrium (NE). Assume there are $N$ players in a game. Let $S_{i}$
and $U_{i}$ for $1\leq i\leq N$ be the strategy spaces and the payoff, or
utility, for each player $i$, respectively. The individual elements of the
strategy space $S_{i}$ for player $i$ are called the \textit{pure strategies}%
. The game can then be described \cite{Maille2010} by the set $G:$

\begin{equation}
G:\left\{ N;S_{1},S_{2},...S_{N};U_{1},U_{2},...U_{N}\right\} .
\label{GameDef}
\end{equation}
The presumable outcome of the game is determined by analyzing the behaviour
of the players and their strategy choices. Let $s=\left\{
s_{1},s_{2},...s_{N}\right\} $ be the profile of pure strategies with $%
s_{i}\in S_{i}$. Let $s_{-1}$ be the profile of strategies excluding player $%
i$. A strategy profile $s$ with $s=(s_{i};s_{-i})$ for all $i$, is a NE \cite%
{Nash1950,Nash1951,Myerson1991,Rasmusen1989,Osborne2004} such that for all $%
1\leq i\leq N$ we have

\begin{equation}
U_{i}(s)\geq U_{i}(t;s_{-i})\text{ for all }t\in S_{i}\text{.}  \label{NE}
\end{equation}
This is also described by stating that the strategy of each player $i$ is a 
\textit{best reply} \cite{Myerson1991,Rasmusen1989,Osborne2004} to the
strategies of other players. In the cybersecurity context, the defenders'
strategy profile that is a NE will consist of a set of defensive strategies,
one on the behalf of each defender, such that the strategy of each defender
is a best reply to the strategies of the attackers.

\subsubsection{Nash equilibrium in mixed strategies}

A \textit{mixed strategy} \cite{Osborne2004} is a linear combination, with
real coefficients, of two or more pure strategies, with their probability
weights summing up to $1.$ This defines the probability distribution $%
\pi_{i}=(\pi_{i,t})_{t\in S_{i}} $ for player $i$ choosing randomly among
the pure strategies $S_{i}$. The expected payoff for the player $i$ is then
given \cite{Maille2010} by

\begin{equation}
\bar{U}_{i}(s_{1},s_{2},...s_{N})=\dsum \limits_{j=1}^{N}\dsum
\limits_{s_{j}\in
S_{j}}U_{i}(s_{1},s_{2},...s_{N})\Pi_{k=1}^{N}\pi_{k,s_{k}}.
\label{ExpectedUtility}
\end{equation}
A set of probability distributions $\pi_{i}=(\pi_{i})_{1\leq i\leq N}$
defines a mixed NE, or a NE in mixed strategies, such that for all $i$ and
any other probability distribution $\bar{\pi}_{i}=(\pi_{i,t})_{t\in S_{i}}$
we have

\begin{gather}
\dsum \limits_{j=1}^{N}\dsum \limits_{s_{j}\in
S_{j}}U_{i}(s_{1},s_{2},...s_{N})\Pi_{k=1}^{N}\pi_{k,s_{k}}\geq  \notag \\
\dsum \limits_{t\in S_{i}}\dsum \limits_{j\neq i}\dsum \limits_{s_{j}\in
S_{j}}U_{i}(t;s_{-i})\bar{\pi}_{i,t}\Pi_{k\neq i}\pi_{k,s_{k}}.
\label{MixedNE}
\end{gather}
A key result of Nash's thesis \cite{Nash1950,Nash1951} states that a NE
always exists in mixed (randomized) strategies in games where each player
has only a finite number of deterministic strategies. In a NE, no one player
can improve his/her situation by unilaterally changing his/her strategy.
This amounts to stating that each person is doing as well as they possibly
can, even if that does not mean that an optimal outcome has been achieved
for the collective of all players. In a cyber attack when a NE is determined
for a team of defenders, neither is left with any motivation to deviate
unilaterally from it.

\subsubsection{Prisoners' Dilemma}

The game of \textit{Prisoners' Dilemma} (PD) \cite%
{Myerson1991,Rasmusen1989,Osborne2004} describes the following situation: a)
Two criminals, called in the following as Alice and Bob, commit a crime
together and are arrested. As the evidence is being investigated, they wait
for their trial, b) Each suspect is offered the opportunity to confess the
crime after placing him/her in a separate cell, c) Each suspect may choose
between the strategies of confessing $(\mathcal{D})$ or not confessing $(%
\mathcal{C})$, where $\mathcal{C}$ and $\mathcal{D}$ represent the pure
strategies of \textit{cooperation} and \textit{defection} with one's partner
in the crime and not with the authorities, d) If neither of the two
confesses, i.e. $(\mathcal{C},\mathcal{C}),$ they both go free, and divide
between them the proceeds of their crime. We represent this in the following
by $3$ units of payoff to each prisoner, e) However, if one prisoner
confesses $(\mathcal{D})$ and the other does not $(\mathcal{C})$, the
prisoner who confesses testifies against his partner in exchange for going
free and gets the entire $5$ units of payoff. However, the prisoner who did
not confess is sent to prison and that is represented by the payoff of zero,
f) If both suspects confess, i.e. $(\mathcal{D},\mathcal{D})$, then both are
convicted while a reduced term is given to both. This is represented by
giving each suspect $1$ unit of payoff. This payoff is better than having
the other suspect confess, but it is not so good as going free. The game
between the prisoners can be represented by the following bimatrix of
payoffs:

\begin{equation}
\begin{array}{c}
\text{Alice}%
\end{array}%
\begin{array}{c}
\mathcal{C} \\ 
\mathcal{D}%
\end{array}
\overset{\overset{%
\begin{array}{c}
\text{Bob}%
\end{array}
}{%
\begin{array}{ccc}
\mathcal{C} &  & \mathcal{D}%
\end{array}
}}{\left( 
\begin{array}{cc}
(3,3) & (0,5) \\ 
(5,0) & (1,1)%
\end{array}
\right) }  \label{PDmatrix1}
\end{equation}
where the first and the second entry in a bracket correspond to Alice's and
Bob's payoff, respectively. Let Alice play $\mathcal{C}$ with probability $p$
and play $\mathcal{D}$ with probability $(1-p)$. Similarly, let Bob play $%
\mathcal{C}$ with probability $q$ and play $\mathcal{D}$ with probability $%
(1-q)$. The players' payoffs for the PD matrix (\ref{PDmatrix1}) are

\begin{equation}
U_{A}(p,q)=-p+4q-pq+1,\text{ }U_{B}(p,q)=4p-q-pq+1.  \label{MixedPayoffsPD}
\end{equation}
The inequalities that define the NE consisting of a pair of mixed strategies 
$(p^{\ast},q^{\ast})$ in PD can then be written as

\begin{align}
U_{A}(p^{\ast },q^{\ast })-U_{A}(p,q^{\ast })& =-(p^{\ast }-p)(1+q^{\ast
})\geq 0,  \notag \\
U_{B}(p^{\ast },q^{\ast })-U_{B}(p^{\ast },q)& =-(q^{\ast }-q)(1+p^{\ast
})\geq 0,
\end{align}%
which produces a unique NE in PD: $p^{\ast }=q^{\ast }=0$. The NE
corresponds to both players playing the pure strategy $\mathcal{D}$. Ref. 
\cite{Kostyuk2013} Utilizes an international relations PD game to present an
explanation of the complexities of cyber intrusions and the way forward for
nation-states to deal with these new exigencies. The PD game is also
discussed in the context of cybersecurity in Ref. \cite{MetcalfCasey2016}.

\subsubsection{Refinements of Nash equilibrium}

Some of the largest problems in security applications come from actions that
cannot be anticipated. This makes using NE problematic as the concept
presumes that the structure of the game, as well as all possible moves, is
common knowledge among the players. Several refinements of the NE have been
introduced \cite{Fudenberg1991,Osborne1994} including \textit{sequential
equilibrium}, \textit{proper equilibrium}, \textit{trembling hand equilibrium%
}, and \textit{rationalizability}.

\subsection{Sequential or Stackelberg games}

A dynamic model of the duopoly game was proposed by Stackelberg (1934) \cite%
{Gibbons1992,Stackelberg1934} in which a leader (or dominant) firm moves
first and in view of the leading firm's move a follower (or subordinate)
firm moves second. For instance, in the early history of US mobile industry,
General Motors played this leadership role against more than one firm such
as Ford and Chrysler who acted as followers. In the sequential game of
duopoly a \textit{Stackelberg equilibrium}\ is obtained using the solution
concept of \textit{backwards-induction outcome} of the game. As a
solution-concept it is stronger than that of NE and refers to sequential
nature of the game. Multiple NE may appear in sequential move games whereas
only one of those is associated with the backwards-induction outcome of the
game.

Consider the following simple three step game, a) Player $1$ chooses an
action $a_{1}$ from the set $A_{1}$ of his strategies, b) Player $2$
observes $a_{1}$ and then chooses an action $a_{2}$ from the set $A_{2}$ of
her strategies, c) Payoffs for the two players are $U_{1}(a_{1},a_{2})$ and $%
U_{2}(a_{1},a_{2})$. It is an example of the dynamic games of complete and
perfect information whose key features are, a) Players take their moves in
sequence, b) All previous moves are known to the players they make a next
move, and c) The players' payoff functions are common knowledge. Given the
action $a_{1}$ is previously chosen, at the second stage of the game when
player $2$ takes his turn to make the move he faces the problem:

\begin{equation}
\underset{a_{2}\in A_{2}}{\mathrm{Max}}U_{2}(a_{1},a_{2}).  \label{max2}
\end{equation}
Assume that for each $a_{1}$ in $A_{1}$, player $2$'s above optimization
problem has a unique solution $R_{2}(a_{1})$, which is the best response of
player $2$. By anticipating player $2$'s response to each action $a_{1}$
that player $1$ might take, Player $1$ can now solve player $2$'s
optimization problem. So that player $1$ faces the problem:

\begin{equation}
\underset{a_{1}\in A_{1}}{\mathrm{Max}}U_{1}(a_{1},R_{2}(a_{1})).
\label{max1}
\end{equation}
Assuming that this optimization problem has a unique solution for player $1$
and it is denoted by $a_{1}^{\star}$. The solution $(a_{1}^{%
\star},R_{2}(a_{1}^{\star}))$ is then called as the \textit{%
backwards-induction outcome} of this game. In Ref. \cite%
{DamjanovicBehrendt2017} Damjanovic-Behrendt presents an approach to
optimize the cybersecurity decisions in order to protect instances of a
federated Internet of Things platform in the cloud. His solution implements
the repeated Stackelberg security game. An overview of use-inspired research
in Stackelberg security games is presented in Ref. \cite{Kar2016}.

\subsection{Repeated games}

A specific class of dynamic games are the repeated games in which the
players play the same game more than once. Players observe the outcome of
the first play before the start of the second play. Payoffs for the entire
game are then obtained as the sum of the payoffs from the previous stages.
Generally, repeated games have a strategic structure that is more complex
than it is in their one-stage counterpart. This is because the players'
strategic choices in the following stages are influenced by the outcome of
the choices they make in an earlier stage.

A two-stage game of complete but imperfect information is sequential in that
the players' moves in the first stage are observed before the next stage
begins. The simultaneity of the players' moves in each stage result in the
imperfect information in the game. Such a game consists of these steps \cite%
{Gibbons1992}, a) Players $A$ and $B$ simultaneously choose their moves $p$
and $q$ from their strategy sets $\mathcal{P}$ and $\mathcal{Q}$,
respectively, b) Players $A$ and $B$ observe outcome of the first stage of
the game, $(p,q)$, and they then simultaneously choose actions $p_{1}$ and $%
q_{1}$ from the sets $\mathcal{P}$ and $\mathcal{Q}$, respectively, c)
Payoffs are $U_{i}(p,q,p_{1},q_{1})$ for $i=A,$ $B$. Usually, the games from
this class are solved using the method of backwards-induction. This involves
solving the simultaneous-move game between players $A$ and $B$ in the second
stage, given the outcome from the first stage. Players $A$ and $B$ can
anticipate that their second-stage behavior will be given by $(p_{1}^{\star
}(p,q),q_{1}^{\star }(p,q))$. In view of this, the first-stage interaction
between the players becomes equivalent to the following simultaneous-move
game: a) Players $A$ and $B$ simultaneously choose actions $p$ and $q$ from
sets $\mathcal{P}$ and $\mathcal{Q}$, respectively, b) Payoffs are $%
U_{i}(p,q,p_{1}^{\star }(p,q),q_{1}^{\star }(p,q))$ for $i=A,B$. When $%
(p^{\star },q^{\star })$ is the unique NE of this simultaneous-move game,
the set of four numbers $(p^{\star },q^{\star },p_{1}^{\star
}(p,q),q_{1}^{\star }(p,q))$ is known as the \textit{subgame-perfect outcome}
\cite{Gibbons1992} of this two-stage game. This solution concept is the
natural analog of the backwards-induction outcome in games of complete and
perfect information.

Consider the PD game given by the matrix (\ref{PDmatrix1}) for which the
players play the game twice and the outcome of the first play is observed
before the second stage begins. Payoffs for the entire game are then
obtained as the sum of the payoffs from the two stages of the game. The game
is a two-stage game of complete but imperfect information \cite{Gibbons1992}%
. Assume players $A$ and $B$ play the pure strategy $C$ with probabilities $%
p $ and $q$, respectively, in stage $1$. Also assume the players $A$ and $B$
play the strategy $C$ with probabilities $p_{1}$ and $q_{1}$, respectively,
in stage $2$. Let $U_{A1}$ and $U_{B1}$ represent the payoffs to players $A$
and $B$, respectively, in the stage $1$. From Eqs. (\ref{MixedPayoffsPD})
these payoffs are $U_{A1}=-pq+4q-p+1$ and\ $U_{B1}=-pq+4p-q+1.$ The NE
conditions for this stage are $U_{A1}(p^{\star },q^{\star
})-U_{A1}(p,q^{\star })\geq 0,$ $U_{B1}(p^{\star },q^{\star
})-U_{B1}(p^{\star },q)\geq 0$ giving $p^{\star }=q^{\star }=0$ (i.e.
defection for both the players) as the unique NE in this stage. Similarly,
in the second stage the payoffs to players $A$ and $B$ are expressed as $%
U_{A2}$ and $U_{B2}$ respectively, where $U_{A2}=-p_{1}q_{1}+4q_{1}-p_{1}+1,$
$U_{B2}=-p_{1}q_{1}+4p_{1}-q_{1}+1$. Therefore, the strategy of defection,
i.e. $p_{1}^{\star }=q_{1}^{\star }=0$, once again comes out as the unique
NE in the second stage. To compute the subgame-perfect outcome of this
two-stage game, we analyze its first stage given that the second-stage
outcome is also the NE of that stage ---namely $p_{1}^{\star }=q_{1}^{\star
}=0$. For this NE the players' payoffs in the second stage are $%
U_{A2}(0,0)=1,$ $U_{B2}(0,0)=1.$ The players' first-stage interaction,
therefore, in this two-stage game becomes equivalent to a one-shot game, in
which the payoff pair $(1,1)$ from the second stage is added to their
first-stage payoff pair. We can write the players' payoffs in the one-shot
game as $U_{A(1+2)}=U_{A1}+U_{A2}(0,0)=-pq+4q-p+2,$ and $%
U_{B(1+2)}=U_{B1}+U_{B2}(0,0)=-pq+4p-q+2.$ It has again $(0,0)$ as the
unique NE. The unique subgame-perfect outcome of the two-stage PD,
therefore, is $(0,0)$ in the first stage, and it is also $(0,0)$ in the
second stage. The strategy of defection in both the stages comes out as
subgame-perfect outcome for the two stage classical PD.

\subsection{Cooperative games}

In cooperative games, players are allowed to form coalitions, binding
agreements, pay compensations, make side payments etc and there is a strong
incentive to work together to receive the largest total payoff. In their
pioneering work on game theory \cite{Neumann1944}, von Neumann and
Morgenstern offered models of coalition formation where the strategy of each
player consists of choosing the coalition s/he wishes to join. In coalition
games the players' possibilities are described by the available resources of
different groups (coalitions) of players and joining a group, or remaining
outside, is part of strategy of a player affecting his/her payoff. The
notion of a strategy disappears in a cooperative game and the notion of a
coalition and the value or worth of that coalition attain significance. It
is assumed that each coalition can guarantee for its members a certain
amount that is called the \textit{value of a coalition }\cite%
{Fudenberg1991,Osborne2004}. It measures the worth of the coalition that is
obtained as the payoff which the coalition can guarantee for itself if it
selects an appropriate strategy. However, the `odd man' can prevent the
coalition from receiving more than this amount.

An example of a three-player symmetric cooperative game is a classical
three-person normal form game \cite{Burger1963} that is defined by:

a) Three non-empty sets $\Sigma_{A}$, $\Sigma_{B}$, and $\Sigma_{C}$ that
are the strategy sets of the players $A$, $B$, and $C$,

b) Three real valued functions $U_{A}$, $U_{B}$, and $U_{C}$ that are
defined on $\Sigma_{A}\times\Sigma_{B}\times\Sigma_{C}$, and

c) The product space $\Sigma_{A}\times\Sigma_{B}\times\Sigma_{C}$ that is
the set of all tuples $(\sigma_{A},\sigma_{B},\sigma_{C})$ with $%
\sigma_{A}\in\Sigma_{A}$, $\sigma_{B}\in\Sigma_{B}$ and $\sigma_{C}\in%
\Sigma_{C}$.

For this game, a strategy is understood as such a tuple $(\sigma_{A},%
\sigma_{B},\sigma_{C})$ and $U_{A}$, $U_{B}$, $U_{C}$ are payoff functions
of the three players and the game can be denoted as $\Gamma=\left\{ \Sigma
_{A},\Sigma_{B},\Sigma_{C};U_{A},U_{B},U_{C}\right\} $.

Let $\Re=\left\{ A,B,C\right\} $ represent the set of players and assume
that $\wp$ is an arbitrary subset of $\Re$. Players in $\wp$ may form a
coalition so that the coalition $\wp$ can be considered as a single player.
It is expected that players in $(\Re-\wp)$ will form an opposing coalition
and the game has two opposing \textquotedblleft coalition
players\textquotedblright\ i.e. $\wp$ and $(\Re-\wp)$. One of the two
strategies $1$, $2$ is chosen by each of the three players $A$, $B$, and $C$%
. There is no payoff if the three players choose the same strategy. If the
two players choose the same strategy, both receive one unit of money from
the 'odd man.' The payoff functions $U_{A}$, $U_{B}$ and $U_{C}$ for players 
$A$, $B$ and $C$, respectively, are given as \cite{Burger1963}:

\begin{align}
U_{A}(1,1,1) & =U_{A}(2,2,2)=0,  \notag \\
U_{A}(1,1,2) & =U_{A}(2,2,1)=U_{A}(1,2,1)=U_{A}(2,1,2)=1,  \notag \\
U_{A}(1,2,2) & =U_{A}(2,1,1)=-2,  \label{PayoffsCoop}
\end{align}
with similar expressions for $U_{B}$ and $U_{C}$. Suppose $\wp=\left\{
B,C\right\} $, hence $\Re-\wp=\left\{ A\right\} $. The coalition game
represented by $\Gamma_{\wp}$ is given by the payoff matrix:

\begin{equation}
\begin{array}{c}
\left[ 11\right] \\ 
\left[ 12\right] \\ 
\left[ 21\right] \\ 
\left[ 22\right]%
\end{array}
\overset{%
\begin{array}{cc}
\left[ 1\right] & \left[ 2\right]%
\end{array}
}{\left( 
\begin{array}{rr}
0 & 2 \\ 
-1 & -1 \\ 
-1 & -1 \\ 
2 & 0%
\end{array}
\right) }.
\end{equation}
Here the strategies $\left[ 12\right] $ and $\left[ 21\right] $ are
dominated by $\left[ 11\right] $ and $\left[ 22\right] $. After eliminating
these dominated strategies the payoff matrix becomes

\begin{equation}
\begin{array}{c}
\left[ 11\right] \\ 
\left[ 22\right]%
\end{array}
\overset{%
\begin{array}{cc}
\left[ 1\right] & \left[ 2\right]%
\end{array}
}{\left( 
\begin{array}{cc}
0 & 2 \\ 
2 & 0%
\end{array}
\right) }.
\end{equation}
It is seen that the mixed strategies:

\begin{equation}
\frac{1}{2}\left[ 11\right] +\frac{1}{2}\left[ 22\right] ,\text{ and }\frac{1%
}{2}\left[ 1\right] +\frac{1}{2}\left[ 2\right] ,  \label{cltCoop}
\end{equation}
are optimal for $\wp$ and $(\Re-\wp)$ respectively. With these strategies a
payoff $1$ for players $\wp$ is assured for all strategies of the opponent;
hence, the value of the coalition $\upsilon(\Gamma_{\wp})$ is $1 $ i.e. $%
\upsilon(\left\{ B,C\right\} )=1$. Since $\Gamma$ is a zero-sum game $%
\upsilon(\Gamma_{\wp})$ can also be used to find $\upsilon(\Gamma_{\Re-\wp})$
as $\upsilon(\left\{ A\right\} )=-1$. The game is symmetric and one can write

\begin{align}
\upsilon(\Gamma_{\wp}) & =1\text{, \ \ and\ \ \ }\upsilon(\Gamma_{\Re-\wp
})=-1\text{ or,}  \notag \\
\upsilon(\left\{ A\right\} ) & =\upsilon(\left\{ B\right\} )=\upsilon
(\left\{ C\right\} )=-1,  \notag \\
\upsilon(\left\{ A,B\right\} ) & =\upsilon(\left\{ B,C\right\}
)=\upsilon(\left\{ C,A\right\} )=1.  \label{VcltC}
\end{align}
Cooperative game theory has been applied to cybersecurity in a number of
studies: In a Masters thesis submitted to the Florida Atlantic University,
Golchubian \cite{Golchubian2017} has used cooperative game theory by
developing a game theoretical approach to prevent collusion and to
incentivize cooperation in cybersecurity contexts. Vakilinia and Sengupta 
\cite{Vakilinia2017} have investigated profit sharing in coalitional game
theory using calculation for rewarding the players that is
participation-fee. In particular, they analyze the well-known \textit{%
Shapley value} concept \cite{Rasmusen1989,Gibbons1992} by formulating a
coalitional game between organizations in cybersecurity information sharing
system.

\subsection{Bayesian games}

In other situations that are characterized by the players' access to only a
partial knowledge about the game, game theory is still shown to be an
effective modelling tool by exploiting the concepts from Bayesian games \cite%
{Gibbons1992}. A Bayesian game is defined as a game of incomplete
information in which the players do not have the complete knowledge of the
rules of the game. The incomplete knowledge is described by the existence of
the so-called state of Nature, which is decided probabilistically by some
relevant random source. In Bayesian games, the probability distribution over
the states of Nature is private to each player and which represents each
player's knowledge about Nature. Nature is allowed to leak some information
about its state in the Bayesian games, which is called the signal to the
players. With the signal, the players can probabilistically work out their
expected utilities. A Bayesian game \cite{Pham2015} consists of a tuple $%
\left\langle N,\Omega ,\left\langle S_{i},T_{i},C_{i},\tau
_{i},p_{i},U_{i}\right\rangle _{i\in N}\right\rangle $ where $\Omega $ is
the set of natural states, and for each player $i\in N,$

a) $S_{i}$ is the set of player $i$'s all available actions,

b) $T_{i}$ is the set of player $i$'s signals/types, with $%
\tau_{i}:\Omega\longrightarrow T_{i}$ is the state-to-signal mapping,

c) $C_{i}:T_{i}\longrightarrow2^{S_{i}}$ is the set of $i$'s available
actions after receiving $t_{i}\in T_{i},$

d) $p_{i}$ is the probability measure over $\Omega,$ and,

e) $U_{i}:\Omega\times S\longrightarrow\mathcal{R}$ is player $i$'s utility
function where $\mathcal{R}$ is the set of real numbers.

The solution concept of a NE is adapted into Bayesian games and is called
Bayesian NE. Some applications of Bayesian games include Liu et al's. \cite%
{Liu2006} computation of Bayesian Nash outcomes for an intrusion detection
game and under the conditions of limited information, Johnson et al's \cite%
{Johnson2010} determination of Bayesian Nash equilibria for network security
games.

\section{Network/cybersecurity and game theory}

The information technology landscape has been revolutionized by the recent
advances in software and hardware technologies. Cyberspace has now become an
integral part of the way the business is conducted. For current
telecommunication and information networks, their network/cybersecurity is
the main concern and the protection and security of cyberspace
infrastructure is of key importance.

Game theory is applied to networks in settings in which agents are connected
by physical or virtual links. Given the network structure and the actions of
other users of the network, the agents must decide on some action in a
strategic manner.

Heterogeneous, large-scale, and dynamic networks define the cyberspace of
the present time. Cyberspace has become increasingly complex even within
carefully designed network and software infrastructures. Ample and a large
attack surface is available for evasive maneuvers of adversaries in the
cyberspace. Cyberspace has become characterized by higher computational
power and ubiquitous connectivity and these features have given birth to new
risks and threats.

The miscreants launching cybersecurity attacks have various degrees of
uncertainty and defenders have incomplete information about their intentions
and capabilities. Improving cybersecurity thus involves difficult challenges
and decision making on multiple levels and over different time scales. The
goal of cybersecurity is to provide practical and scalable security
mechanisms and to enhance the trustworthiness of cyber-physical systems.

As is the case with the physical security, in cybersecurity there exists a
wide variety of the agents' utilities, including adversarial and
antithetical types. Game theory, therefore, shares many common features with
the cybersecurity problem. The success of a cybersecurity scheme depends not
only on the actual cyberdefense strategies that have been implemented, but
also on the strategic actions taken by the attackers to launch their
attacks. Thus these scenarios are well-suited to the game theoretical
analyses of the cybersecurity schemes. Such analyses can also be viewed from
the perspective of establishing trust. When security is compromised,
building trustworthy relationships, and deciding whether to trust received
information becomes particularly relevant. It is well known \cite%
{Witteloostuijn2003} that the trust problem can be formulated as in
game-theoretic strategic terms. Trust emerges as an important aspect in the
design and analysis of security solutions and the implementations of
security games involve several levels of trust.

\subsection{Network/cybersecurity games}

A significant motivation for cybersecurity games comes from earlier
applications of game theory to the domain of physical security. These are
examples of practical situations that demonstrate the potential for game
theory in that domain. Physical security considerations are important at
airports, product transportation, national security patrols, etc. Usually, a
defender allocates the available resources to defend against an attacker
whereas the attacker can attempt to compromise targets that the defender is
protecting from possible attacks. Most often, the defender can best allocate
resources to minimize the chance of success for the attacker and minimize
the cost incurred by the defender. How should the defender allocate agents,
patrols, surveillance technology, and other resources to minimize the impact
of attackers? Examples of physical security situations include, a) the
airport security: where the defender can schedule optimal checkpoints and
patrols for their agents, b) the coast guard: more efficiently protection
can be provided to ferries or ports that are the targets for theft or
terrorism. The finite number of agents and limited resources can be
allocated in such ways to best counteract wide scale poaching.

In network/cybersecurity situations, the zero-sum games between malicious
attackers and the transmitter-receiver pairs can model the problems of
jamming and eavesdropping in communication networks. Attackers and defenders
are most often considered as the agents in network security problems.
Security games form a basis for formal decision making, algorithm
development, and in predicting the behaviour of attackers\textit{. }Security
games can be deterministic or stochastic. They can be sequential or
hierarchical (Stackelberg game) in which an agent has a certain information
advantage over the others. In cooperative or coalitional security games the
agents can cooperate to achieve their strategic objectives. Examples of
security games in the network/cybersecurity domain include, i) intrusion
detection \cite{Alpcan2003,Dritsoula2012}, ii) privacy concerns \cite%
{Manshaei2013,Rajtmajer2017,Raya2010}, iii) network jamming \cite%
{Basar1983,Altman2009,Ai2017}, and iv) eavesdropping in communication
networks \cite{Han2009}.

Scheduling and deployment of patrols is a key operational problem for those
who are responsible for the security of airports, art galleries etc. Alpern 
\textit{et al.} \cite{Alpern2011} have presented a class of patrolling games
addressing the optimization problem involving randomized, and thus
unpredictable, patrols. They have considered the facility to be patrolled as
a network or graph \textsl{Q} of interconnected nodes (e.g. rooms,
terminals) such that the Attacker has the option to attack any node of 
\textsl{Q} within a given time \textsl{T}. That is, the attacker requires 
\textsl{m} consecutive periods that are uninterrupted by the Patroller in
order to commit his nefarious act and therefore win. In this approach, the
Patroller can follow any path on the graph. The patrolling game turns out to
be a win-lose game in which, given best play on both sides, the \textit{Value%
} is the probability that the Patroller successfully intercepts an attack.

\subsection{Examples\label{examples}}

We begin by reviewing two examples from the literature in some detail, as
reported by Sokri \cite{Sokri2018} and Durkota et al~\cite{Durkota2017}.

\subsubsection{Optimal resource allocation in cybersecurity}

Sokri \cite{Sokri2018} has considered a security game between an attacker $a$
and a defender $d$ in a system for cyberinfrastructure. Let $T=\left\{
t_{1},t_{2},...,t_{n}\right\} $ be a set of $n$ targets that are at the risk
of being attacked and $S=\left\{ s_{1},s_{2},...,s_{m}\right\} $ a set of
resources to protect the targets. Vector $\left\langle a_{t}\right\rangle $
can represent the attacker's mixed strategy where $a_{t}$ is the probability
of attacking the target $t$. The defender's mixed strategy is the vector $%
\left\langle p_{t}\right\rangle $ where the marginal probability of
protecting the target $t$ is $p_{t}$. Players' access to mixed strategies
allows them to play probability distributions over their pure strategies. A
strategy profile $\left\langle a,p\right\rangle $ is a combination of
(mixed) strategies that the attacker and the defender may play. Let $%
r_{d}(t) $ be the defender's reward if the attacked target $t$ is covered
and $c_{d}(t)$ his cost if the target is uncovered. Similarly, denote by $%
r_{a}(t) $ the attacker's reward if the attacked target $t$ is uncovered and
by $c_{a}(t)$ the attacker's costs if the attacked target $t$ is covered.
For the strategy profile $\left\langle a,p\right\rangle $ following are the
expected payoffs of the two players:

\begin{align}
U_{d}(a,p) & =\dsum \limits_{t\in T}a_{t}\left[
p_{t}r_{d}(t)-(1-p_{t})c_{d}(t)\right] ,  \label{DefenderPayoff} \\
U_{a}(a,p) & =\dsum \limits_{t\in T}a_{t}\left[
(1-p_{t})r_{a}(t)-p_{t}c_{a}(t)\right] .  \label{AttackerPayoff}
\end{align}
The payoffs in Eqs. (\ref{DefenderPayoff},\ref{AttackerPayoff}) depend only
on the attacked targets and their protection and these payoffs do not
consider the targets that are not attacked. Now, if the players move
simultaneously, the solution of this cybersecurity game is a NE. However, if
the game is played sequentially in which the defender moves first (leader)
and commits to a strategy and the attacker (follower) reacts to the
defender's move, the Stackelberg equilibrium appears as the standard
solution in this leader-follower interaction.

Given the defender's strategy $p$, the attacker's optimization problem can
be presented as follows:

\begin{gather}
\mathrm{Max}_{a}\dsum \limits_{t\in T}a_{t}\left[
(1-p_{t})r_{a}(t)-p_{t}c_{a}(t)\right] , \\
\text{s. t. }\dsum \limits_{t\in T}a_{t}=1,\text{ }a_{t}\geq0,\text{ }%
\forall t\in T.
\end{gather}
It is optimal to assign $1$ to any $a_{t}$ that is associated with a maximal
value of

\begin{equation}
U_{a}(t,p)=(1-p_{t})r_{a}(t)-p_{t}c_{a}(t),\text{ }\forall t\in T.
\end{equation}
The dual problem that corresponds to the above has the same optimal solution
and it can be formulated as follows:

\begin{gather}
\mathrm{Min}\text{ }u, \\
u\geq U_{a}(t,p),\text{ }\forall t\in T.
\end{gather}
The complementary slackness condition then becomes:

\begin{equation}
a_{t}(u-U_{a}(t,p))=0,\text{ }\forall t\in T.
\end{equation}
When the leader problem is completed by including the follower's optimality
condition, it becomes a single \textit{mixed-integer quadratic problem} \cite%
{Coniglio2014}:

\begin{gather}
\mathrm{Max}_{p}\dsum \limits_{t\in T}a_{t}\left[
p_{t}r_{d}(t)-(1-p_{t})c_{d}(t)\right] ,  \label{1st} \\
\sum_{t\in T}p_{t}\leq m,  \label{2nd} \\
\dsum \limits_{t\in T}a_{t}=1,  \label{3rd} \\
0\leq u-U_{a}(t,p)\leq(1-a_{t})M,\text{ }\forall t\in T,  \label{4th} \\
p_{t}\in\lbrack0,1],\text{ }\forall t\in T,  \label{5th} \\
a_{t}\geq0,\text{ }\forall t\in T,  \label{6th} \\
u\in\mathcal{R}.  \label{7th}
\end{gather}
Eq. (\ref{1st}) maximizes the leader's expected payoff. The coverage to the
available resources $(m)$ is limited by Eq. (\ref{2nd}) whereas Eq. (\ref%
{5th}) restricts the coverage vector to $[0,1]$. The leader's mixed strategy
is enforced to be feasible by these two constraints. Eq. (\ref{4th}), where $%
M$ is a large number, is the complementary slackness condition indicating
that the follower's payoff $u$ is optimal for every pure strategy with $%
a_{t}>0.$

Sokri \cite{Sokri2018} has considered the example of a game in normal form
as shown in the Table 1 and that is adapted from the Refs. \cite%
{Jain2010,An2011}. There are $4$ targets and two resources that can cover
any of the two targets. For each target, there are two payoffs i.e. the
payoffs of the attacker and the payoffs of the defender. Each payoff
consists of two parts i.e. a reward and a cost.

\begin{gather*}
\overset{%
\begin{array}{ccccccc}
&  &  & \text{Defender} &  & \text{Attacker} & 
\end{array}
}{%
\begin{tabular}{lcccc}
& Reward & Cost & Reward & Cost \\ 
Target 1 & $4$ & $3$ & $9$ & $6$ \\ 
Target 2 & $3$ & $2$ & $7$ & $6$ \\ 
Target 3 & $6$ & $4$ & $10$ & $8$ \\ 
Target 4 & $3$ & $2$ & $12$ & $6$%
\end{tabular}
} \\
\text{Table 1: Payoff table \cite{Sokri2018}.}
\end{gather*}
Note that, a) If the target is attacked, the defender can cover a target and
get a reward, b) He can also leave the target uncovered and incur a cost if
it is attacked, c) If the target is uncovered, the attacker can attack a
target and get a reward, and d) If the target is covered he can also incur a
cost. By changing the static values in Table 1 to a range of values, an
uncertainty can be placed on each variable. Using a three-point estimate
(minimum, most likely, and maximum) approach that incorporates this
uncertainty, Sokri \cite{Sokri2018} has determined the following solution,
which is found to satisfy all the constraints as well as the numerical
convergence criterion:

\begin{equation}
\left\langle p=(0.5549,\text{ }0.4994,\text{ }0.3411,\text{ }0.6025),\text{
\ \ }a=(0,\text{ }0,\text{ }0,\text{ }1)\right\rangle .
\end{equation}
The objective did not move significantly after many iterations, and even if
it is heavily defended the attacker preferred to attack the most valuable
target. The most likely payoffs have the corresponding cumulative
distribution function (CDF). This can now be determined With this solution
and the median of the defender's average payoff comes out to be
approximately 0.95. This gives a 50\% probability that the defender's
average payoff will be less than 0.95. The values for minimum and maximum of
defender's average payoff are then determined to be 0.4261 and 1.5166,
respectively.

\subsubsection{Threshold-setting to detect data exfiltration}

Data breach involves strategic interaction between defender and attacker for
which game theory provides helpful insights. It is carried out through the
process of information exfiltration and involves unauthorized transfer of
information. A dynamic (sequential) game model of data infiltration is
described by Durkota et al~\cite{Durkota2017} in which the attacker's
objective is to exfiltrate as much data as possible before the activity is
detected. The defender's objective is to minimize the loss of data before
the breach is detected.

The defender records the volume of data that each host at the network
uploads over time while using windows of time with fixed lengths. Defender
selects a detection threshold $\theta $, chosen from a set $\Theta $ of
thresholds, such that if the host uploading data that is more than $\theta $
in the time window then it triggers an alarm.

The defender can set the detection threshold $\theta $ for each host
individually. However, it is possible to identify groups of hosts with
similar behaviours. For instance, a group can be of type $\lambda $ from the
set $\Lambda $ of all types. For a randomly selected host, $P(\lambda )$
then defines the probability that the host is of the type $\lambda $. That
is, $P(\lambda )$ is the probability of the concurrence of the host types.
It is assumed that both the attacker and the defender know the probability $%
P(\lambda )$. Two hosts of same types have the common activity pattern i.e. $%
P(o\mid \lambda )$ gives the probability that a host of type $\lambda $
transfers the amount of data $o\in O$ in a time interval.

It can be the case that even without an attacker's activity a selected
threshold $\theta $ is surpassed along with the alarm triggered. These
instances are called the false positives and usually it is a time consuming
task for the administrators to determine their cause. Certain number of
false positives are expected in the defender's strategies and usually their
bound is expressed as the constant FP.

The external and internal attackers are called the outsider and the insider,
respectively. The nature of information that the insiders and the outsiders
have about the targeted organization can be different from each other.
Although the outsider may know which host types exist but cannot know which
types were compromised. In contrast, the insider knows which host types
exist and also which were compromised.

\textbf{Defender's Strategy:} Defender's pure strategy $\psi $ is a map from
the set $\Lambda $ of all types to the set $\Theta $ of thresholds, i.e. $%
\psi :\Lambda \rightarrow \Theta $. Defender's mixed or randomized strategy
is $\sigma (\theta \mid \lambda )$ that defines a probability distribution
of thresholds $\theta $ given host types $\lambda $. The false positive
constraint for a defender strategy $\sigma $ is then written as

\begin{equation}
\sum\limits_{\lambda \in \Lambda }\sum\limits_{\theta \in \Theta }\sigma
(\theta \mid \lambda )P(\lambda )\text{FP}(\theta \mid \lambda )\leq \text{FP%
},
\end{equation}%
where

\begin{equation}
\text{FP}(\theta \mid \lambda )=\sum\limits_{o\in O:\text{ }o>\theta
}P(o\mid \lambda ),
\end{equation}%
is type $\lambda $'s amount of false positives when the threshold is $\theta 
$.

\textbf{Attacker's Strategy:} It consists of choosing the amount of data $%
a\in A$ that the attacker infiltrates in the next time window. By
controlling one of the users, the attacker uploads as much data as possible
before being detected. The attacker is awarded a utility when the sum of the
host's activity $o\in O$ and the amount of data that the attacker
infiltrates $a\in A$ is less than the detection threshold $\theta \in \Theta 
$. A belief state

\begin{equation}
b\in \Delta (\Lambda \times \Theta )
\end{equation}%
is a probability distribution over possible host types and threshold
settings. It is assumed that first the defender selects the threshold $%
\theta $ and the attacker acts in response to knowing $\theta $. Attacker's
expected utility is then defined as $u_{a}(\sigma ,\pi )$ where $\sigma $ is
the defender's mixed strategy whereas $\pi $ is the attacker's policy chosen
from the set $\Pi $. The attacker's policy

\begin{equation}
\pi :\Delta (\Lambda \times \Theta )\rightarrow A,
\end{equation}%
is defined as a mapping from the set of belief states $\Delta (\Lambda
\times \Theta )$ to the set $A$ of the amounts of data. During the course of
interaction with the defender, the attacker takes into account the last
action and observation and uses Bayesian update rule in order to keep track
of his belief $b$. Attacker's action $\pi $ depends on his belief $b$ i.e. $%
\pi =\pi (b)$. The defender's expected utility is defined as

\begin{equation}
u_{d}(\sigma ,\pi )=-Cu_{a}(\sigma ,\pi ),
\end{equation}%
where $C>0$. This requires that the attacker and the defender have opposite
objectives and their payoffs are proportional to each other's. Also, $C$
being greater than $1$ means that the defender's disutility is greater than
the attacker's utility.

\textbf{The insider vs. outsider attacks:} The data breach attacks can be
performed by agents who are inside an organization/company or who are
inside. The outsiders usually does not know the type of the host that is
compromised even though s/he can know which types (group) exist within the
company, for instance, IT admins and secretaries.\ The insiders, however,
know their host types as they use the network regularly and they are also
knowledgeable on the defences that are deployed. For instance, an insider
because of him/her knowing the exact values of the thresholds that has been
fixed for each host type by the defender, can exfiltrate exactly at those
values.

An approximate algorithm is used to compute the attacker's policy and to
find an approximate Stackelberg equilibrium. The defender's strategy is also
an approximate Stackelberg equilibrium and his utility presents as a close
approximation to the exact Stackelberg equilibrium.

\textbf{Defender's optimal strategy against attacks by the insiders:}
Durkota et al~\cite{Durkota2017} present an algorithm that computes exact
Stackelberg algorithm against attacks from the insiders. Knowing the type of
the user, using whom the insider can exfiltrate data, allows representing
the game between the host and the attacker as a normal form game. For this
game, the attacker's strategy consists of choosing for each host type a
probability distribution over the actions from the set $A$. Similarly, the
defender's strategy consists of choosing for each host type a probability
distribution over thresholds from the set $\Theta $. The game between the
attacker and all host types can then be formalized as one problem. To
achieve this, the zero-sum normal-form linear program \cite{Shoham2008} is
extended to include a false-positive constraint and multiple host types:

\begin{gather}
\min_{\sigma (\theta \mid \lambda )}U_{a},  \label{cond1} \\
\text{s.t. : }(\forall \lambda \in \Lambda ,\text{ }\forall a\in A):\text{ }%
\sum\limits_{\theta \in \Theta }u_{a}(\theta ,a,\lambda )\sigma (\theta \mid
\lambda )\leq U_{a,\lambda },  \label{cond2} \\
\sum\limits_{\lambda \in \Lambda }P(\lambda )U_{a,\lambda }\leq U_{a},
\label{cond3} \\
(\forall \lambda \in \Lambda ):\sum\limits_{\lambda \in \Lambda }\sigma
(\theta \mid \lambda )=1,  \label{cond4} \\
(\forall \lambda \in \Lambda ,\text{ }\forall a\in A):\sigma (\theta \mid
\lambda )\geq 0,  \label{cond5} \\
\sum\limits_{\lambda \in \Lambda }\sum\limits_{\theta \in \Theta }P(\lambda
)\sigma (\theta \mid \lambda )\text{FP}(\theta \mid \lambda )\leq \text{FP}.
\label{cond6}
\end{gather}%
Here $\sigma (\theta \mid \lambda ),U_{a}$ and $U_{a,\lambda }$ are the
variables in the linear program. In the above, the expected utilities of
each type are $U_{a,\lambda }$ that is weighed by its probability given by (%
\ref{cond3}). With the requirement (\ref{cond1}), the expected utility of
the attacker $U_{a}$ is minimized. With the requirement (\ref{cond2}) it is
ensured that against the given defense strategy, a best response is played
in each host type. The requirements given by (\ref{cond4}) and (\ref{cond5})
are placed in order to ensure that the defender's strategy given by $\sigma $
is considered a proper probability distribution. The requirement (\ref{cond6}%
) ensures that the false-positive rate is met by $\sigma $.

\textbf{Defender's optimal strategy against attacks by the outsiders:} As
the outsider is unaware of the host's type, s/he tries to learn about it by
observing host's activity. This results in the attacker's strategies
becoming more complex when these are compared to the strategies of the
insider. To achieve his objectives, the outsider can come up with stronger
attacks which can vary over time. The uncertainty involved suggests using
Partially Observable Markov Decisions Processes (POMDPs). The algorithms
developed in order to solve POMDPs can be used to compute the attacker's
best response and his/her optimal strategy. In every time step, the attacker
exploiting the POMDP framework takes an action from a set of allowed actions
and receives the environment's response to that action. Based on this
response he then updates his beliefs about the environment. Also, in each
time step, the attacker's utility is a function of his action and the
environment's response. POMDP generates a solution in the form of a policy
describing the list of actions for all belief states about the environment.
Heuristic Search Value Iteration (HSVI) \cite{Smith2004} is the well
established algorithm that is used to solve the POMDPs as it computes the
Stackelberg equilibrium of the game. Using interations the HSVI computes the
strategies constituting the best response of the attacker and the defender.
A best response is then achieved by a convergence of the best response
strategies.

\section{Application scenarios}

Game theory has many security applications, and we cannot give detailed
examples of all such analyses here. In this section, we review a number of
other applications that have seen game-theoretic analysis.

\subsection{Trust assignment}

Rajtmajer~\emph{et~al.}~(2017)~\cite{Rajtmajer2017} consider the problem of
multiparty access control~\cite{Hu2012}. Users of social networks have a
shared interest in the privacy settings applied to content relating to them.
They model the problem as a variant of the ultimatum game where all parties
are motivated to reach some agreement, despite the need to compromise. They
develop a model of how participants will vary their offers over time, and
show empirically that the network tends to converge around the proposals of
the more `stubborn' users who are unwilling to vary their proposals.
However, they also find that stubborn users are less likely to reach
agreement with their neighbours at all, unlike less stubborn users who will
quickly take an approach similar to that of their neighbours, resulting in a
greater rate of successful interaction.

Raya~\emph{et~al.}~(2010)~\cite{Raya2010} consider the \textquotedblleft
free-rider\textquotedblright\ problem in systems based on data aggregation:
participants gain a privacy benefit by refusing to trust other parties with
their data, but with less data available, the system as a whole becomes less
resistant to malicious behavior. They show that it is possible to design an
incentive scheme that discourages free-riding to avoid a `tragedy of the
commons'-type scenario.

\subsection{Resource allocation}

Game theory is a natural tool for the analysis of resource allocation
problems in cybersecurity. An example of this is the analysis by Panaousis~%
\emph{et~al.}~(2014)~\cite{Panaousis2014}, which builds up a quantitative
model of how various security controls interact with various classes of
vulnerability, yielding different types of costs to the defender, for
example in the form of reputational damage or data loss. This is then used
to argue that certain controls are or are not worthwhile at a given budget
and at a given depth in the network.

A related analysis is given by Cui~\emph{et~al.}~(2017)~\cite{Cui2017}. In~%
\cite{Cui2017}, it is hypothesized that an attacker can choose between
attacking a customer database and attacking individual users. Gaining access
to the database yields a greater reward for the attacker, but may lead to a
more vigorous law-enforcement response. Conversely, targeting individual
users---e.g.\ by phishing---leads to a reduced payoff, but may be less risky
for the attacker. The defender is represented by two parties: a system
administrator who manages the database, and a user who sets a security level
for themselves only. One of the more interesting features of this model is
the use of a two-stage process for compromising the database, in order to
model the greater technical sophistication of such an attack, at least
relative to the difficulty of acquiring user credentials by e.g.\ phishing.

\subsection{Anomaly detection}

Intrusion detection systems based on anomaly detection~\cite{Wang2004}
require the setting of a threshold parameter, that determines whether some
data is reported as `normal' or `anomalous'. This leads to a trade-off: a
low threshold will force attackers to sacrifice the efficacy of their
attacks in order to stay covert, but will also lead to a high false-positive
rate, resulting in excess cost to the defender. Conversely, a high threshold
will reduce the time wasted investigating false-positives, but allows
attackers to be less covert and use more powerful attacks that are more
costly to the defender.

This interplay between the strategies of the attacker and defender is
well-modelled by game-theory, and so game-theoretic methods can effectively
inform the design of these anomaly detectors.

Schlenker~\cite{Schlenker2017} considers the problem of allocating
investigative resources to security-relevant events in a more general sense.
A system that triggers investigation in too-predictable a manner is
vulnerable to an attacker that can tailor its behavior so as to avoid a
follow-up investigation even if it is detected. For example, a system that
directs all its investigative capacity toward targets labelled as high-value
is easily circumvented by an attacker who has \emph{carte blanche} to attack
moderate-value systems without concern for covertness.

\subsection{Information flow}

Durkota \emph{et~al.}~(2017)~\cite{Durkota2017} consider the problem of
detecting data exfiltration in a heterogeneous network. Once an attacker is
present inside a network, they must decide how quickly to exfiltrate the
data that they acquire: a small flow of data is difficult to detect, but the
value of the information to the attacker is less timely and therefore argued
to be less valuable. Conversely, a large flow of data is more readily
apparent, but more valuable to the adversary while it goes undetected. This
leads to an interesting result: the optimum strategy for the defender is to
vary their detection threshold randomly, yielding a 30\% reduction in
exfiltrated data relative to a deterministic choice of threshold.

Alvim \emph{et~al.}~(2017)~\cite{Alvim2017} consider information leakage in
more general terms, defining a framework of \emph{information leakage games}%
, and finding that in many cases, the \emph{attacker} also benefits from a
mixed strategy. They also show that the utility of a strategy for the
defender is a convex function, allowing the optimal strategy to be
determined using normal optimization techniques.

\subsection{Deception}

Others have used game theory to model techniques aimed at \emph{deception}
of attackers~\cite{Cohen1998}. Underbrink (2016)~\cite{Underbrink2016}
classifies these into \emph{passive methods}, which serve to frustrate
reconnaissance and detect the attacker before it strikes, and \emph{active
methods}, which in which the defender takes actions predicted to interfere
with an attack in progress.

Schlenker~\cite{Schlenker2018} considers the passive case where the defender
manipulates their behavior so that an attacker scanning the defender's
network will be uncertain about the type or value of each system, making it
difficult for the attacker to effectively allocate their effort. Schlenker
shows that determining the optimal strategy for the defender is NP-hard in
general, but provides an algorithm to approximate this.

Alternatively, deception may be used to engage an attacker that has already
compromised the defender's network. Hor\'{a}k \emph{et al.} (2017)~\cite%
{Horak2017} consider a system in which the defender can feed the attacker
with useless data once an attack has been detected. They argue that evicting
the attacker immediately upon detection is suboptimal, as this leads to the
attacker starting again from an `undetected' state, only now armed with
useful information on the defender's detection capabilities. The defender
might therefore be better served by allowing the attacker to remain for a
time, ideally fed with a stream of valueless disinformation.

\subsection{Jamming}

Game-theoretic analysis of channel jamming has a long history, tracing back
to Basar~(1983)~\cite{Basar1983}, who shows that the optimum strategy of a
single attacker seeking to jam a Gaussian memoryless channel is to either
transmit a linear function of the transmitted signal or to transmit random
symbols, depending on the relative signal and noise powers.

Other authors have carried out similar analyses in different situations: for
example Kashyap \emph{et al.} (2004) \cite{Kashyap2004} consider a Rayleigh
fading channel, and show that knowledge of the channel input does not affect
the jammer's strategy.

Altman \emph{et al.} (2009) \cite{Altman2009} analyze the case of \emph{%
multiple} attackers who seek to jam an orthogonal frequency-division
multiplexing (OFDM) communication channel. The attacker and defender must
each decide how to distribute their power across the available subchannels
in order to minimize or maximise, respectively, the
signal-to-interference-plus-noise ratio (SINR) of the channel.

Han \emph{et al.} (2009) \cite{Han2009} consider a different scenario in
which `friendly' jammers broadcast their own signals, introducing noise to
disrupt eavesdroppers. They consider the problem from an economic viewpoint:
what price can the friendly jammers demand for their services, given some
desired rate of secret communication? However, like all economic analyses,
this depends strongly on the model of the participants: their analysis
assumes that the sender gains a constant utility per unit bandwidth, and the
jammers pay a constant amount per unit power. Nevertheless, the results are
interesting: in their simulations, they find that there exists a cutoff
price for jamming power above which the use of friendly jammers is no longer
justifiable.

\subsection{Smart grids}

Smart grids can incorporate fine-grained demand-side data into their control
systems, as well as provide demand-side management: with the right
incentives, users will consent to automatic reduction of their power
consumption at times of high load---for example, by slightly increasing the
target temperature of their air conditioners, or by delaying the activation
of refrigerator motors.

By incorporating incentives into the pricing scheme, users may be
incentivized to lie about their usage in order to secure a reduced tariff.
Mohsenian-Rad \emph{et al.} (2010)~\cite{MohsenianRad2010} provide a
game-theoretic analysis of a decentralized demand-side management system;
they show how to design the system so that users do not benefit from lying
to each other about their usage.

Though a decentralized system as in~\cite{MohsenianRad2010} might provide
privacy benefits to its users, issues such as communication complexity and
deployment considerations may result in a centralized system being
preferable in practice. Hajj and Awad~(2015)~\cite{HajjAwad2015} describe a
centralized system that uses game-theoretic methods to provides optimum
scheduling. This comes at the cost of forcing users to reveal their
projected demand to the supplier. In practice this may be a reasonable
sacrifice: in order to take advantage of off-peak tariffs, users must
already reveal some information on their demand schedule, so the difference
in privacy might well be small in practice.

\section{Challenges to applying game theory to security}

Although game theory has been shown to be significant for security, there
exist many challenges that need to be addressed for developing a viable
game-theoretic approaches to security. In this regard, some key challenges
include the complexity of computing a game-theoretical equilibrium strategy,
as the illustrative examples in Section (\ref{examples}) show. There are
also difficulties in properly quantifying security parameters such as risk,
privacy, and trust \cite{Rajtmajer2017}\cite{Raya2010}, i.e. the parameters
in terms of which the utility functions for the participants (players) in a
security game are defined.

Choosing an appropriate game model for a given security problem comes out as
a challenge for the game theory too. Such a model need to depend on the
detail and particular aspects of the security problem/application scenario.
Choosing a game can be solely based on the intuition and this choice may not
substantiated by the available data. A two-player game can be a model for a
security game involving an attacker and a defender. However, in the dynamic
version of this game can involve multiple stages for attacking and
defending. In fact, as described in Section (\ref{examples}), the games of
later type are more likely to be representative of the network/cybersecurity
challenges of the real world.

Another aspect of the security game models is that the players are assumed
to have unbounded rationality. In real life and experimental studies, the
players do not always act with rationality. As a consequence, there exists a
significant scope for studying the solution-concepts of Harsanyi's disturbed
games or that of Selten's perturbed games \cite{Osborne2004}\cite%
{Harsanyi1988} in the network/cybersecurity situations. In Selten's
perturbed games, a player's hand `trembles', resulting in the erroneous move
and the trembles are assumed to be determined by a random process. On the
other hand, in Harsanyi's disturbed games, it is the payoffs or the utility
functions, rather than the players' actions, that go astray.

Interpretation of game-theoretical notions such as mixed strategy Nash
equilibrium also appears as a challenge, and particularly so for the
security games. Usual approach in game theory in this regard involves
considering repeated games whereas many security games are represented as
one-shot games. Even within the game theory community, there is no consensus
on how to interpret a mixed strategy. There is clear need for interpreting
the notion of a mixed-strategy for network/cybersecurity games. In order to
convert the game theoretic results into practical security solutions these
challenges are required to be addressed.

\section{Acknowledgement}

Mingyu Guo's and M. Ali Babar's contributions were partially funded by the
Cyber Security Research Centre Limited whose activities are partially funded
by the Australian Government's Cooperative Research Centres Programme.

\end{document}